# Reference Phase of Fresnel Zone Plates

G. W. Webb

Institute for Pure and Applied Physical Sciences, University of California, San Diego

San Diego, CA 92093

The standard zone plate assumes that the shortest ray connecting a radiation source and a detection point has a phase of 0° thereby defining a reference phase. Here we examine the experimental consequences of varying this reference phase from 0° to 360°. It is concluded that reference phase is an intrinsic and useful property of zone plates.

## I. Introduction

Fresnel zone plates are a type of diffractive lens. They are used to focus radiation in a broad range of physical areas. These areas include the focusing of electromagnetic radiation from the microwave band [1,2] to x-rays [3] to de Broglie matter waves [4]. In addition to such Fresnel zone plates focusing in the spatial domain, analogues have been created for focusing in the spectral domain [5] and the time domain [6]. Here we consider the role of phase in spatially focusing Fresnel zone plates, in particular the assumption of a type of reference phase which is inherent to the zone plate.

The concept of a focusing diffractive lens grew out of Fresnel's analysis of the diffraction of radiation through an aperture [7]. Fresnel analyzed radiation emitted from a source that passed through an aperture and arrived at a detection point. He called attention to zones of constructively and destructively interfering radiation in the aperture. In defining these zones, Fresnel assigned a phase origin or reference phase of $0°$ to *the shortest ray* connecting the source and detection point in order to simplify the analysis. Based on Fresnel's analysis, Soret later demonstrated focusing in half opaque zone plates which blocked radiation from what was defined as destructively interfering zones [8]; the same choice of phase origin as Fresnel was used. Soret also demonstrated the existence of focusing by blocking the "constructive" zones instead. Later work has shown that the out-of-phase zones need not be blocked if selective phase shifts are introduced [1,2,10]. However the same special choice of phase origin as Fresnel has been made. To the best of our knowledge the consequences of this specific choice of phase have not been explored or are not widely known.

Here we explore the effects of the choice of phase with experimental measurements using planar Fresnel zone plates of the blocking type, but without loss of generality. Fig. 1a shows rays from a source (S) passing through an aperture (A) in an opaque screen to a detection point (P). Rays have a phase at P which depends on the positions of S and of P, the distance between them, and on the point where they went through the aperture. Time dependence is suppressed. The path length of the ray connecting S and P, sometimes called the direct ray, is R=f+h. Historically the phase of a general ray with path length r=r1+r2 has been computed by subtracting R from r, implicitly assuming that the distance R defines the phase origin $\theta_{ref}=0°$ or reference phase. We use a definition of phase generalized to include $\theta_{ref}$ explicitly

$$\text{phase} = (r1 + r2 - R)\frac{360°}{\lambda} - \theta_{ref} \qquad (1)$$

where $\lambda$ is the wave length. Fig. 1b assumes $\theta_{ref}=0°$ and shows the relative phase at P of rays plotted with a gray scale on the plane of the aperture at the point where the ray passed through the aperture. However if a different reference phase is chosen, for example $\theta_{ref}=60°$ in Fig. 1c, then the relative phase distribution across the aperture is changed. Fig. 1d displays the sine of the phase plotted vs. radius for the two choices $\theta_{ref}=0°$ and $\theta_{ref}=60°$; 0<sin(phase)<1 corresponds to gray to white, while -1<sine(phase)<0 corresponds to black to gray in Figs. 1b,c. All sin(phase)>0 rays can be taken as in-phase and sin(phase)<0 rays as out-of-phase. A Fresnel zone plate is designed to have a geometry that blocks the out-of-phase rays at P so that the only rays arriving are in-phase and thus a beam of radiation is focused at P.

## II. Experimental Zone Plates.

We computed the required planar Fresnel zone plate geometry for a variety of different choices for $\theta_{ref}$ [11]. In Fig. 2 they are shown numbered ZP=0-11 (ZP#) with corresponding $\theta_{ref}$=0°-330° in increments of 30°. It can be seen that zone plate features change smoothly with ZP# or equivalently $\theta_{ref}$. The specific parameters are those of Fig. 1. This choice of parameters gives an equal number of in-phase and out-of-phase zones for $\theta_{ref}$=0° such that the total area of in-phase zones does not change significantly as $\theta_{ref}$ is varied. In this context note that in the series ZP0-5, as the center constructive zone becomes progressively more destructive, the outermost destructive zone becomes more constructive.

The amplitude $|U_{calc}|$ and phase $\theta_{calc}$ of the focused radiation at P can be calculated in the scalar approximation of Fresnel - Kirchhoff [12]:

$$U_P = -\frac{ike^{-i\omega t}}{4\pi} \iint_A U_0 \frac{e^{ik(r1+r2)}}{r1r2}[\cos(n,r1) - \cos(n,r2)]dA \qquad (2)$$

$U_P$ is the "optical disturbance" at P, n is the normal to the aperture, and the other quantities have the same meaning as Fig. 1a. Eq. 2 was first solved for ZP=0-11 assuming a constant (uniform) feed pattern, $U_0$, for simplicity. The results $|U_{calc}|$ and $\theta_{calc}$ are plotted in Fig. 3 vs. ZP# and its corresponding $\theta_{ref}$. Time dependence is suppressed. Note that $|U_{calc}|$ is nearly constant across ZP0-11 while $\theta_{calc}$ varies nearly linearly with ZP# and $\theta_{ref}$.

Based on the calculations in Fig.2, a series of experimental zone plates were fabricated using printed circuit board materials and techniques. They consisted of 15 μ (ca) thick copper rings on a low-loss dielectric substrate of 168μ thickness [13]. These zone plates were attached to flat, low-loss, low refractive index dielectric foam substrates of 6.3mm thickness. Fabrication techniques and materials were the same for all zone plates.

**III. Zone Plate Measurements and Comparison with Model**

The zone plates were measured in the apparatus of Fig. 4a,b. The feed (F) is an open ended waveguide positioned behind the zone plate at the design focusing point. The power $P_{beam}$ reaching the detector (D) was measured for each zone plate under identical conditions in Fig.4a. Beam amplitude, $|U|$, was defined through $P_{beam}=|U|^2$. The phase apparatus in 4b has arms defined, by F and horn H2, which are connected by a -10dB directional coupler (DC). Relative amplitude between arms is adjusted by attenuator (A) and relative phase (θ) by translating horn H2 toward H1 under micrometer control.

Fig. 5 displays measured beam amplitude data $|U|$ for ZP0-11 as points. The feed pattern of F was determined in an independent measurement of F and parameterized; this parameterized function was included as $U_0$ in the calculation of Eq.2. The solid line shows resultant calculated values for $|U|$. Note that the calculation qualitatively reproduces the magnitude of the overall variation across ZP0-11 (about 1.7dB in power), the existence of a slight maximum in amplitude for ZP=11, and the minimum near ZP=6.

The phase at P of the focused beam, $\theta_{beam}$, was measured with the phase apparatus of Fig. 4b. In these measurements, ZP=0 was mounted first and θ and A adjusted for null; θ was adjusted by the position of H2. Then ZP1-11 were installed, A adjusted, and H2 moved under micrometer control for null. The change in the null position of H2 from ZP0 was converted to phase for ZP1-11 using the known λ. Thus measured phase is relative to that of ZP0. These measured beam phase data, $\theta_{beam}$, are plotted vs. $\theta_{ref}$ as points in Fig. 6. The calculated phase of the focused beam was also determined from Eq. 2 using the parameterized feed function and is shown in Fig. 6 by the solid line. Note that measured and calculated phases both vary close to linearly with $\theta_{ref}$.

It is also possible to fix the phase in the *upper arm* of the interferometer of Fig. 4b and independently vary the phase of the *lower arm* through the ZP#. Figure 7a-c shows detector power for three different fixed positions of the upper arm which are equivalent to: a) $\theta_{ref}=0°$ (ZP=0), b) $\theta_{ref}=180°$ (ZP=6) c) $\theta_{ref}=270°$ (ZP=9). The points are the power at D of the interferometer vs. $\theta_{ref}$ as ZP# is varied. Since the amplitude and phase of each zone plate were independently measured above, the output of the interferometer could also be predicted as a $\theta_{ref}$, as shown by the solid curves in Figure 7a-c. The agreement between the measurements and predictions is good and displays a smooth self-consistent variation of zone plate properties as $\theta_{ref}$ is varied.

## IV. Discussion and Conclusions

This work has shown that reference phase $\theta_{ref}$, relative to the direct ray, is an intrinsic and useful property of a Fresnel zone plate. It has the important property that it controls the path length through constructive zones in the zone plate and thus the phase of the focused beam. In particular, the beam phase $\theta_{beam}$, varies close to linearly with $\theta_{ref}$. It is inferred that the small deviations from the predicted almost-linear relationship between $\theta_{beam}$ and $\theta_{ref}$ of a zone plate arise here primarily from a departure of the feed function from uniform. Conversely, it has been shown that the design choice for $\theta_{ref}$ from 0° to 330° does not strongly affect the amplitude of the focused beam; it is inferred that the small variations in amplitude that are observed are also due primarily to deviations the experimental feed function from uniform. If the geometry of the zone plate can varied in real-time [14] then the phase of the focused beam can also be controlled in real-time. Finally, since Fresnel zone plates can exist in a variety of arbitrarily curved shapes [1] as well as planar, it seems likely that the phase of focused beams diffracted from arbitrarily curved shapes can also be controlled through the reference phase.

It is a pleasure to thank Susan Angello, Wayne Vernon, Pete Schmid, and P.K. Park for discussions, Brian Maple for providing hospitality and space, and Roger Isaacson for the loan of essential equipment. Initial stage of research was supported under DARPA contract no. DAAH-98-C-R060.

**Figure Captions**

**Fig. 1.** a) Schematic of rays which leave source (S) pass through circular aperture (A) in opaque screen (SC) and arrive at detection point (P), not to scale. S, A, and P are co-axial. R=f+h is the shortest path length between S and P. Path length R defines what is sometimes called the direct ray. Path length of a general ray is r=r1+r2; b) Example gray scale plot of phase at P of rays going through different parts of the aperture with reference phase $\theta_{ref}=0°$ and c) $\theta_{ref}=60°$; d) Sine of the phase from Eq. 1 for $\theta_{ref}=0°$ (solid) and $\theta_{ref}=60°$ (dots) plotted along a radius from center to edge of aperture. 0<sin(phase)<1 corresponds to gray to white and -1<sin(phase)<0 corresponds to black to gray in b,c. Specific parameters used here and below are: Frequency $\nu$=39GHz, f=7.04cm, aperture radius=7.91cm, and R=3.05m.

**Fig. 2.** Zone plates ZP=0, 1, 2, …11, corresponding to reference ray choice of $\theta_{ref}=0°$, 30°, 60°, …330°, as indicated and to scale. Transparent zones are shown as white and opaque zones as black. The set ZP=0-5 and ZP=6-11 are mirror images of each other which differ in $\theta_{ref}$ by 180°. ZP=0 and its inverse ZP=6 correspond qualitatively to the two used by Soret.

**Fig. 3.** Calculated amplitude $|U_{calc}|$ from Eq. 2 (diamonds) and resultant phase at P, $\theta_{calc}$, (squares) for ZP0-11. The solid line is a guide for the eye and the dashed line is a straight line connecting 0° to 360°. $|U_{calc}|$ and $\theta_{calc}$ are plotted vs. ZP# and corresponding reference phase $\theta_{ref}$.

**Fig. 4.** Schematic of apparatus (not to scale) to measure (a) amplitude and (b) phase at P. Source S emits linearly polarized 39GHz radiation through horn ($H_1$), zone plate (ZP) focuses incident radiation on feed (F); distance from $H_1$ to F is the same in both a) and b). Interferometer of (b) is nulled through the relative phase adjustment ($\theta$) in upper arm and attenuator (A) adjustment in lower arm.

**Fig. 5.** Beam amplitude |U| at P as a function of $\theta_{ref}$. Points are measured data and solid line is calculated from Eq. 2 using measured feed function. The dashed line shows the calculated results with a uniform feed function for comparison.

**Fig. 6.** Beam phase at P vs. reference phase $\theta_{ref}$ for the set ZP0-11 relative to ZP0. Points are measured data with the apparatus of Fig. 4b and the solid line is the calculated phase from Eq. 2 with the same feed function of Fig. 5. Error bars set by width of null.

**Fig. 7.** Interferometer measurements for different fixed settings of A and $\theta$ in Fig. 4b. A and $\theta$ were adjusted to null output at D for a) $\theta_{ref}=0°$ (ZP=0), b) $\theta_{ref}=180°$ (ZP=6) c) $\theta_{ref}=270°$ (ZP=9). Points are measured detector power and the solid line is the calculated fit assuming the measured amplitude of Fig. 5 and measured phase of Fig. 6 for ZP0-11. Measured depth of null –59 dBm is limited by dynamic range in the apparatus.

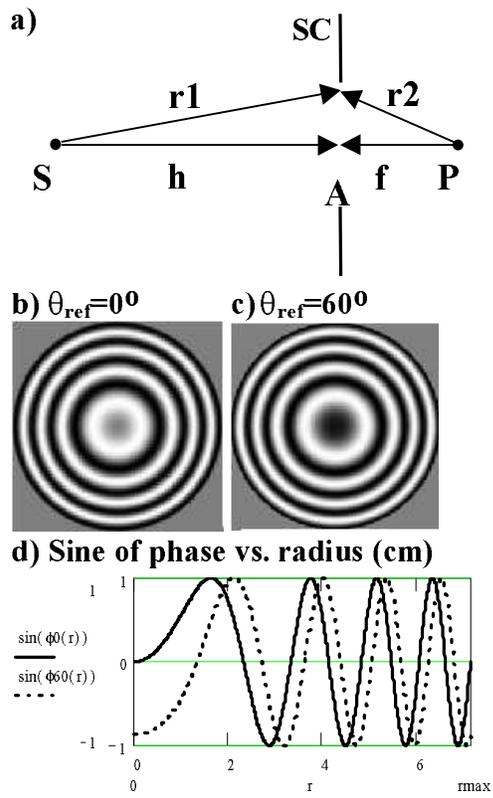

**Fig. 1.**

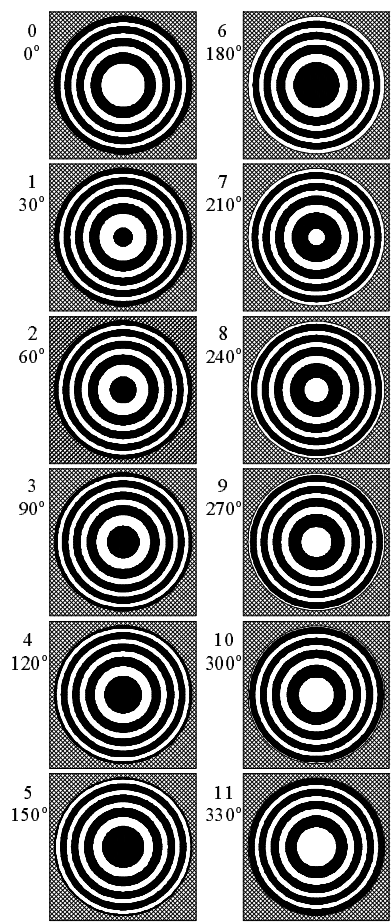

**Fig. 2.**

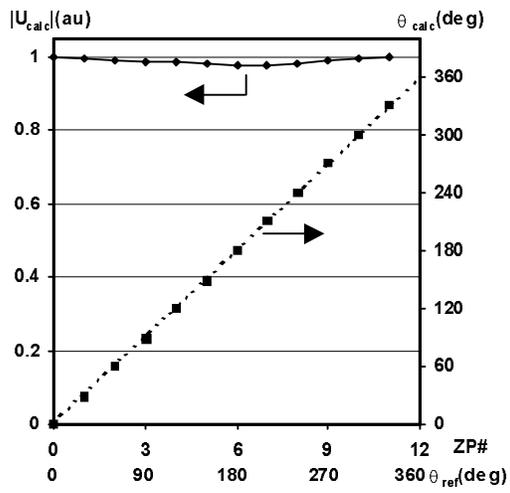

**Fig. 3.**

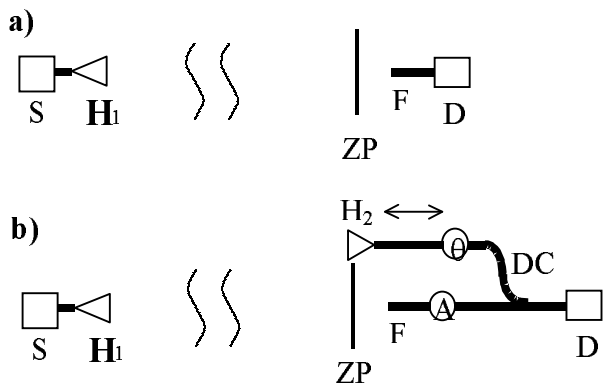

**Fig. 4.**

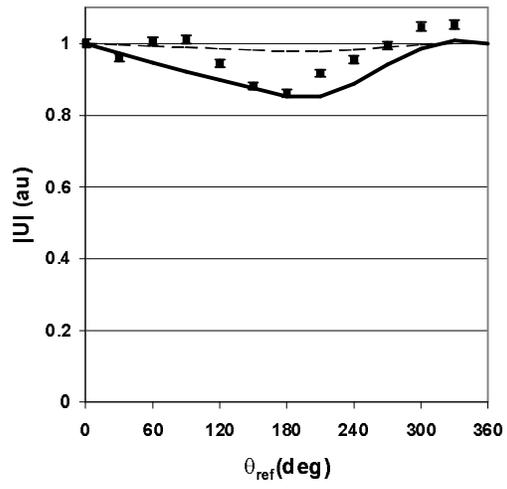

**Fig. 5.**

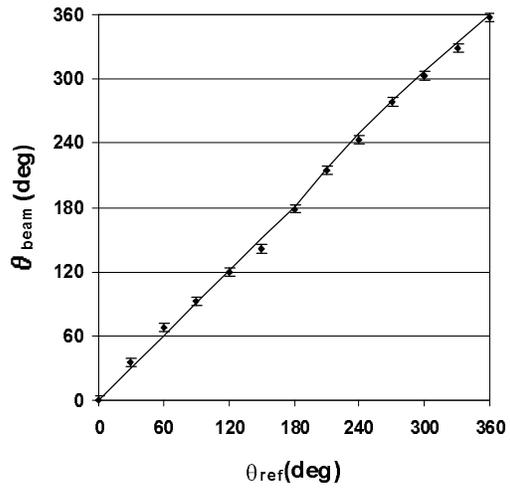

**Fig. 6.**

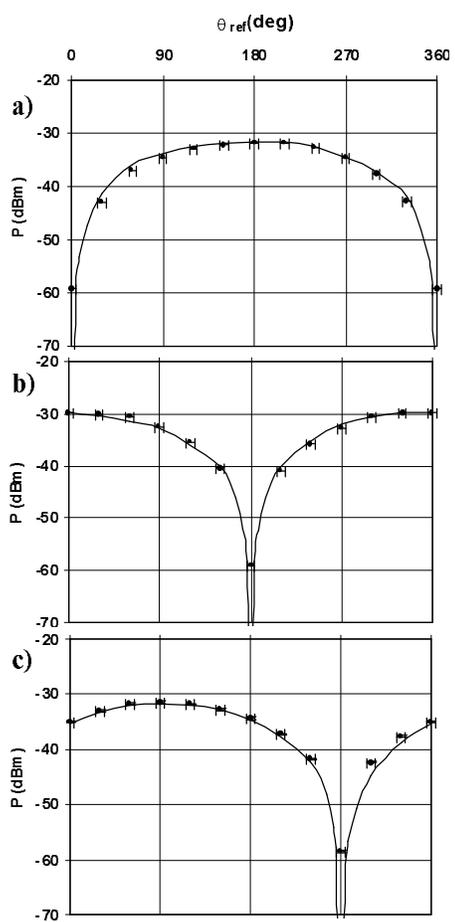

**Fig. 7.**